\documentclass[twocolumn,prd,amsmath,amssymb, aps,superscriptaddress, nofootinbib]{revtex4-1}
\usepackage[version=4]{mhchem} 
\usepackage[normalem]{ulem}
\usepackage{amsmath}
\usepackage{amssymb}
\usepackage{amsfonts}
\usepackage{mathrsfs}
\usepackage{graphicx}
\usepackage{xcolor}
\usepackage{xfrac}
\usepackage{footmisc}
\usepackage{pifont}
\usepackage{times}
\usepackage{epstopdf}
\usepackage{enumerate}
\usepackage[hidelinks]{hyperref}
\usepackage{enumitem}
\usepackage{slashed}
\usepackage{relsize}
\usepackage{epsfig}
\usepackage{verbatim}
\usepackage{bm}
\usepackage[utf8]{inputenc}
\usepackage{graphics}
\usepackage{subfigure}
\usepackage[english]{babel}
\usepackage{orcidlink}

\definecolor{zima_blue}{HTML}{1393C1}
\hypersetup{setpagesize=false,
bookmarksnumbered=true,
bookmarksopen=true, 
colorlinks=true,
linkcolor=zima_blue,
urlcolor=zima_blue,
citecolor=zima_blue,
linktocpage=false}

\begin{document}

\title{Probing Ultralight Primordial Black Hole Dark Matter with XMM Telescopes}

\author{Jun Guo}
\thanks{Corresponding author: \href{mailto:jguo\_dm@jxnu.edu.cn}{jguo\_dm@jxnu.edu.cn}}
\affiliation{College of Physics and Communication Electronics,
Jiangxi Normal University, Nanchang 330022, China}

\author{Qiang Yuan}
\thanks{Corresponding author: \href{mailto:yuanq@pmo.ac.cn}{yuanq@pmo.ac.cn}}
\affiliation{Key Laboratory of Dark Matter and Space Astronomy, Purple Mountain Observatory,
Chinese Academy of Sciences, Nanjing 210023, China}
\affiliation{School of Astronomy and Space Science, University of Science and Technology of China, 
Hefei 230026, China}

\author{Bin Zhu}

\thanks{Corresponding author: \href{mailto:zhubin@mail.nankai.edu.cn}{zhubin@mail.nankai.edu.cn}}
\affiliation{Department of Physics, Yantai University, Yantai 264005, China}

\begin{abstract}
Primordial black holes (PBHs), originating from the gravitational collapse of large overdensities in the early Universe, emerge as a compelling dark matter (DM) candidate across a broad mass range. Of particular interest are ultra-light PBHs with masses around $10^{14}$ to $10^{17}$ g, which are typically probed by searching their evaporation products. Using the soft X-ray signal measured by the XMM telescopes, we derive constraints on the fraction of PBHs dark matter with masses in the range $10^{15}$-$10^{16}$ g. We find that  observations exclude fraction $f>10^{-6}$ at 95\% C.L. for mass $M_{\rm PBH}=10^{15}$ g.
\end{abstract}

\maketitle

\section{Introduction}

The enigmatic nature of dark matter, which accounts for a substantial portion of the Universe's mass budget, remains a profound puzzle that continues to enthrall the fields of astrophysics and particle physics.~\cite{Bertone:2004pz}. Among the proposed candidates, primordial black hole (PBH)~\cite{Zeldovich:1967lct} is one of the compelling alternatives, which offers a natural explanation for the existence of dark matter~\cite{Chapline:1975ojl}. These black holes originate from high-density fluctuations in the aftermath of inflation~\cite{Carr:1993aq, Ivanov:1994pa, Garcia-Bellido:1996mdl, Randall:1995dj}. With a diverse range of masses that account for various dark matter observations and their non-baryonic nature aligning with gravity-dominated interactions, PBHs exhibit strong gravitational effects that address astrophysical puzzles, including gravitational lensing~\cite{Niikura:2019kqi, Mediavilla:2017bok}.

In particular, the existence of ultra-light PBHs, comparable in mass to asteroids, with lifetimes spanning hundreds to millions of times the age of the universe, presents a fascinating avenue for unraveling the cosmological dark matter enigma. To detect and study these primordial black holes, a plethora of methods and experiments have been proposed, aiming to illuminate the properties, abundance, and implications of PBHs and provide valuable insights into the nature of dark matter. These include microlensing surveys~\cite{Macho:2000nvd,Griest:2013aaa,Niikura:2017zjd,Petac:2022rio}, investigations of cosmic microwave background anisotropies~\cite{Ali-Haimoud:2016mbv,Poulter:2019ooo}, evaporations on big bang nucleosynthesis and the extragalactic photon background~\cite{Carr:2009jm,Lehmann:2018ejc}, analysis of gamma-ray radiation~\cite{Coogan:2020tuf,Capanema:2021hnm,Chen:2021ngo,Agashe:2022jgk,Xie:2023cwi,Korwar:2023kpy}, detection of PBH-generated gravitational waves~\cite{Jung:2017flg,Chen:2019irf,LIGOScientific:2019kan,Kavanagh:2018ggo,Chen:2019xse,Wang:2019kzb,Dror:2019twh}, and examination of astrophysics surveys~\cite{Capela:2013yf,Graham:2015apa,Lu:2019ktw}.

Among these approaches, gamma-ray detection plays a pivotal role, as it arises from the emission of gamma-ray radiation through Hawking radiation --- a consequence of quantum effects near the event horizon of a black hole. Due to their small mass and high curvature resulting from formation in the early universe, PBHs release high-energy particles, including gamma rays. While the emission of Hawking radiation from individual PBHs is exceedingly faint, the collective emissions from a population of PBHs hold the potential for detection as a coherent gamma-ray signal. 
%To capture gamma rays originating from PBHs, specialized instruments and observatories such as NASA's Fermi Gamma-ray Space Telescope and the upcoming e-ASTROGAM mission have been designed, enabling detection and investigation of gamma-ray emissions across a broad energy spectrum. 

In ~\cite{Coogan:2020tuf, Korwar:2023kpy}, the abundance of primordial black holes (PBHs) have been constrained by using COMPTEL data of the gamma-ray from the galactic center in the energy range of $0.7-27$ MeV. The analysis considers not only the photons emitted by evaporating PBHs but also includes secondary particles like pions and charged particles. The COMPTEL data provides the most stringent constraints on PBH masses ranging from near $10^{16} \rm g$ to $10^{18} \rm g$, {leaving a gap in $10^{15}-10^{16} \rm g$ as a stable dark matter}. They also analyze the discovery potential of future MeV gamma-ray telescopes, such as  AMEGO~\cite{Kierans:2020otl}, e-ASTROGAM~\cite{e-ASTROGAM:2017pxr}, and GECCO~\cite{Orlando:2021get}. In~\cite{Saha:2021pqf}, the authors utilize a measurement similar to EDGES~\cite{Bowman:2018yin} to study the global 21 cm signal and constrain the fraction of primordial black holes (PBHs). They analyze non-spinning and spinning PBHs in the mass range of $10^{15}\ \rm g$ to $10^{17}\ \rm g$. The low-mass PBHs significantly impact the observed global 21 cm signal due to the heating of the intergalactic medium by the particles emitted during PBH evaporation. However, the SARAS 3 experiment gives conflict results with the EDGES signal at the $95.3\%$ confidence level~\cite{Singh:2021mxo}.

In this study, we focus on ultra-light PBHs with masses on the order of $10^{15}$ grams. Observations suggest that these entities may play a significant role in the dark matter content. To explore this possibility, we utilize archival data from the XMM instrument and examine the soft X-ray emissions resulting from the evaporation process of PBHs through inverse Compton scattering. In the work of \cite{Mukhopadhyay:2021puu}, the authors investigate the Synchrotron and Inverse Compton signals from PBHs in the mass range of $10^{14}\ \rm g$ to $10^{17}\ \rm g$. They utilize experimental sensitivity from MHz experiments and find that the low-energy scenario with $E_{e^\pm}\sim\ \rm MeV$ generates a significant inverse Compton photon signal around a frequency of $10^5\ \rm MHz$, which is different from the XMM-Newton typical frequency.

\section{X-Ray from PBH evaporation}

To accurately model the photon spectrum arising from the evaporation of PBHs, it is essential to consider the effects of inverse Compton scattering (ICS). Inverse Compton scattering involves the interaction between high-energy electrons and photons, resulting in the transfer of energy to the electrons and a modified photon spectrum. Here, we summarize our approach in the supplementary material~\ref{supp:X-ray} to include ICS in the computation of the photon spectrum.

\begin{equation}
\frac{\mathrm{d} \Phi_\gamma^{\mathrm{tot}}}{\mathrm{d} E_\gamma}=\frac{\mathrm{d} \Phi^{\mathrm{PBH}}_\gamma}{\mathrm{d} E_\gamma}+\frac{d \Phi_{\gamma}^{\mathrm{IC}}}{d E_\gamma}
\end{equation}

The generation of inverse Compton scattered X-ray photons occurs through the upscattering of low-energy photons in the galaxy by positrons emitted from primordial black holes (PBHs). The low-energy photon population in the galaxy encompasses three distinct components: cosmic microwave background (CMB), dust-rescattered infrared light (IR), and optical starlight (SL). Describing their distribution as functions of position and wavelength, denoted as $n_i(\lambda, \vec r)$, we extract this information from the GalProp.

\begin{figure}[htbp]
\centering
\includegraphics[width=0.5\textwidth]{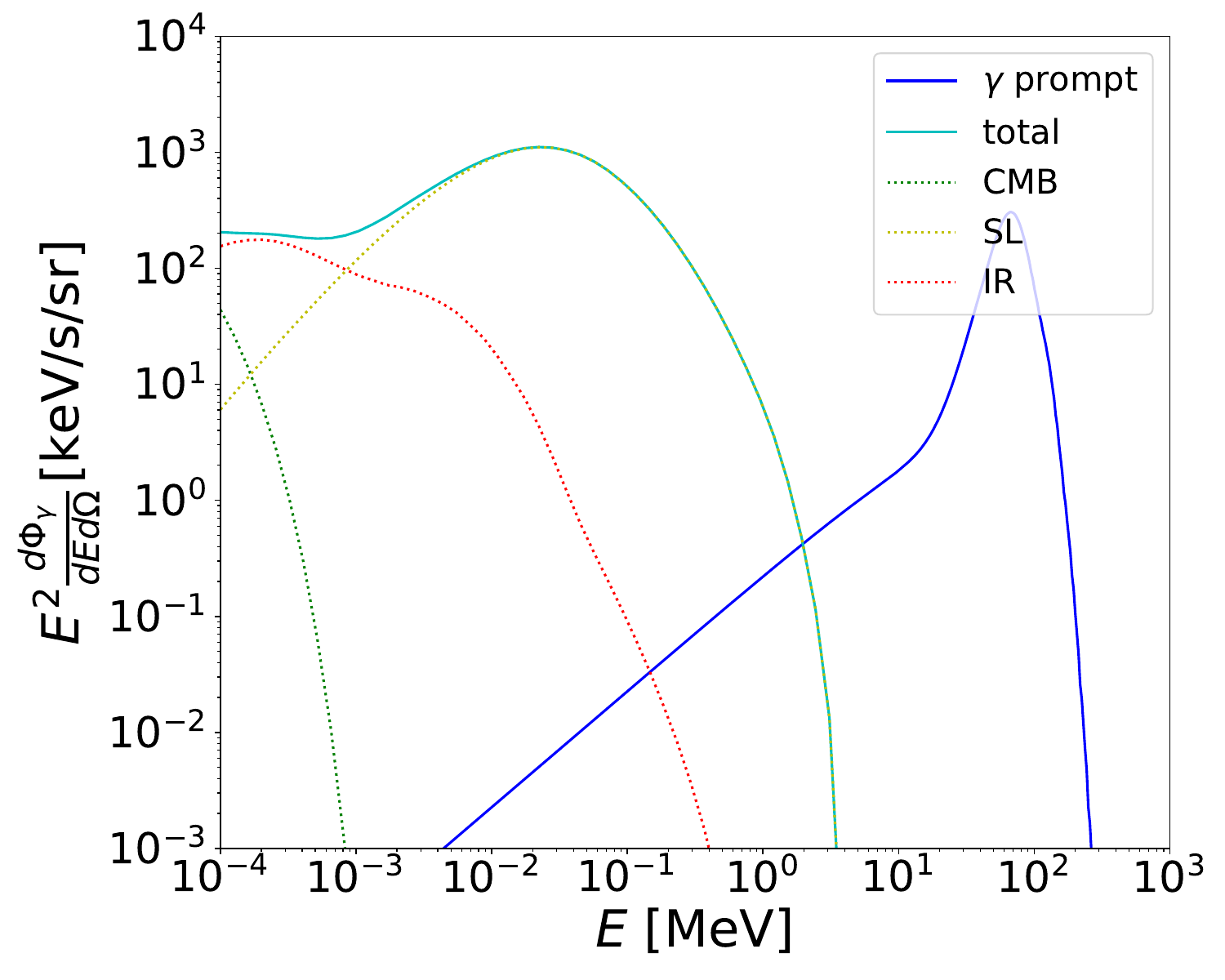}\\
f\caption{The photon flux obtained through inverse Compton scattering (ICS) from primordial black holes (PBHs) with $m_{\rm PBH}=1\times 10^{15}$ g, along with the flux of direct prompt photons.} %Right: The ICS photon flux generate by a heavy PBH with $1\times 10^{16}$ is greatly suppressed.}
\label{Fig:ICS_result}
\end{figure}

In Fig.\ref{Fig:ICS_result}, we present the differential flux generated by the primordial black hole (PBH) for both ICS X-ray and direct prompt gamma-ray emissions. This specific case considers $(l, b) = (10^{\circ},10^{\circ})$ with $m_{\rm PBH} = 1\times 10^{15}$ g and $f_{\rm PBH} = 1$. The ICS signal dominates in the X-ray signal region, while the direct prompt photon signal is concentrated in the MeV scale region. The total ICS photon signal is contributed by three components (CMB, SL, IR). The low-energy X-ray signal ($E_{\gamma} \leq 1$ keV) is mainly attributed to low-energy photons from synchrotron radiation (SL), while the X-ray signal with $E_{\gamma}\simeq 100$ keV is predominantly contributed by infrared (IR) photons. On the right panel, we display the distribution of the same signals for $m_{\rm PBH} = 1\times 10^{15}$ g and $m_{\rm PBH} = 5\times 10^{15}$ g. Both the direct prompt and ICS signals are suppressed in the case of a large $m_{\rm PBH}$.

{The primary characteristic of the inverse Compton (IC) signal is its dominance in the keV energy range. According to Eq.~\ref{Eq:phbspec}, the spectrum of the evaporated electrons is predominantly distributed in the energy range $E_{\gamma}\leq 8\pi T_{\rm BH}\simeq (10^{15}\ \rm gm/M_{\rm BH}) 250\ \rm MeV$. Thus, heavier black holes correspond to lower energies of the evaporated electrons. The evaporated electrons will scatter off the diffuse photons in the Galaxy, which have energies in the eV range. When a MeV-scale electron scatters with a low-energy eV-scale photon, the photon energy in the final state increases  with 
\begin{equation}
E_{\gamma}\simeq 4 E_{\gamma 0}\frac{E_e}{m_e}
\end{equation}
where $E_{\gamma 0}$($E_{\gamma }$) is the initial(final) state photon energy and $E_e$ is the initial state electron energy. For a photon with 1 keV initial energy and $M=10^{15}\ \rm gm$, the final state photon is estimated around 2 keV.}

\section{Methodology and Results}
The dominant inverse Compton scattering (ICS) signal for primordial black holes (PBHs) occurs in the keV range, as shown in Fig.~\ref{Fig:ICS_result}. Light PBHs produce significant X-ray signals, but the signal decreases considerably as the PBH mass increases. To determine the constraint on the PBH fraction $f_{\rm PBH}$, we utilize observations from the XMM-Newton blank-sky survey, as it provides the strictest constraints on X-ray signals.

\subsection{ICS Sigmal Treatment}
We utilize the data provided in~\cite{Foster:2021ngm}. The authors divide the entire sky into $30$ concentric rings centered around galaxy centers, with each ring having a width of $6$ degrees. Each ring contains multiple observations that meet the authors' criteria. Each ring contains the events past the cuts, and the observation area is not the whole ring but these exposures. Each observation $(i)$ includes its solid angle $\Delta \Omega_i$, observed time $t_i$, and observed count $N_{ij}$ at energy bin $E_j$. For comparing the data with the theoretical prediction, we calculate the signal for each ring as follows:

\begin{equation}
N(E) = \sum_i T_i\int_{E'_{min}}^{E'_{max}} R(E, E') dE'\frac{d\phi_{\gamma}^{\mathrm{IC}}}{dE'd\Omega} \Delta\Omega_i
\label{Eq:sig_th}
\end{equation}

Here, $ R(E, E')$ represents the instrument response function. As the experimental observation is provided in a binned format, the integration of energies can be transformed into a summation over bins, {as shown in Eq.~\ref{response_matrix}}. {Show the transformation into summation in supp.} The response matrix between bins is available in the repository~\cite{Foster:2021ngm}. The summation in Eq.~\ref{Eq:sig_th} runs over all the observations within a specific ring. We employ the Navarro-Frenk-White (NFW) profile for calculating the ICS emission in our model, which is expressed as
\begin{equation}
\rho_{\rm NFW}(r) = \frac{\rho_s}{(r/r_s)(r/r_s + 1)^2}
\end{equation}
with $r = \sqrt{s^2+R^2-2sR\cos{l}\cos{b}}$ is the distance to the Galactic center, we set $R=8.33$ kpc and $\rho_S=184$ $\rm MeV/cm^3$.

\subsection{XMM-Newton Data Analysis}

The first $8$ rings are considered as the signal region, while rings $20$ to $31$ are designated as the background region. The background strength is determined by averaging the binned counting rate per keV across the last $11$ rings. The strength of ring $i$ in bin $k$ is calculated as $N_i/(T_i \Delta E_k)$. The signal strength is obtained by subtracting the background strength from the binned counting rate per keV in the first $8$ rings. The uncertainty of the event count is determined using the Poisson distribution rate uncertainty, given by $\sqrt{\rm N}$.

\begin{figure}[ht]
\centering
\includegraphics[width=8cm,height=6cm]{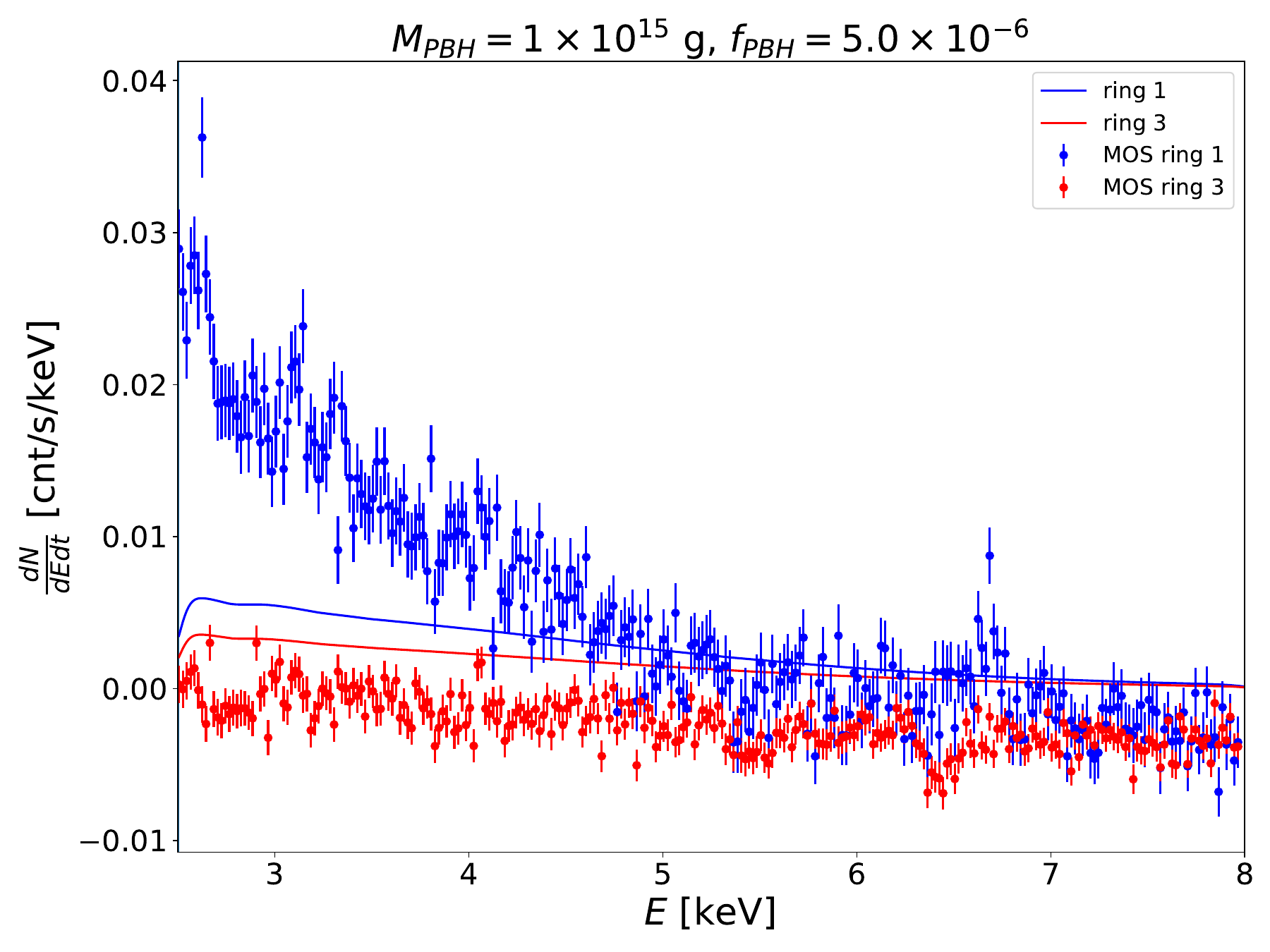}
\caption{We show the predicted signal and data for a light PBH with $m_{\rm PBH} = 1\times 10^{15}$ g and $f_{\rm PBH} = 5\times 10^{-6}$. To show the data of XMM observation clearly, we sampled them. The MOS camera data will give a relatively strict constraint on the fraction parameter $f_{\rm PBH}$.}
\label{Fig:mos_data}
\end{figure}

Fig.\ref{Fig:mos_data} illustrates the ICS signal and experimental data for the rings. The XMM-Newton data is obtained using two separate cameras: MOS and PN, which we analyze independently. The signal region for the MOS data is chosen as 2.5 keV to 8 keV, while for the PN data, it is 2.5 keV to 7 keV. We observe that the PN data provides a more relaxed constraint on the fraction compared to the MOS data, and the inner rings are more tightly constrained than the outer ring. These findings align with the conclusions drawn in Refs.~\cite{Cirelli:2023tnx, DelaTorreLuque:2023nhh}. Based on these findings, we employ the MOS data to constrain $f_{\rm BH}$ and consider the inner 8 rings data as our signal region. For the PN camera data, we ignore it for its low sensitivity. Fig.~\ref{Fig:ring_result} displays some signal and data examples for the MOS camera, clearly demonstrating that the MOS data provides a stringent constraint on $f_{\rm PBH}$. 

\subsection{Likelihood Analysis}
\begin{figure}
\centering
\includegraphics[scale=0.4]{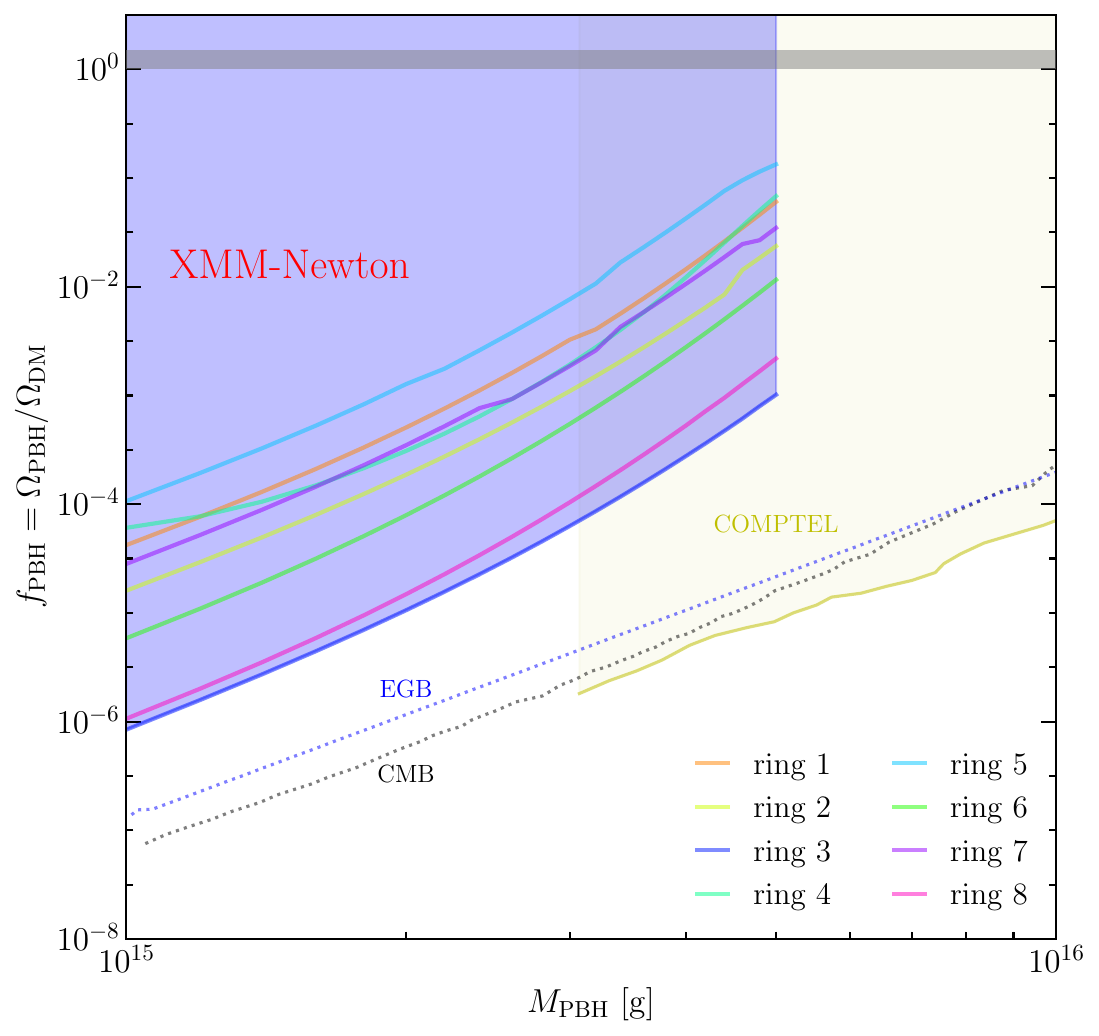}
\caption{The limits on the fraction of PHBs dark matter and their masses from each single ring independently are shown. The most stringent bound is from ring 3. The results of EGRB and COMPTEL are given as well.}
\label{Fig:ring_result}
\end{figure}

We follow the procedures in the work ~\cite{Foster:2021ngm} by using the hypothesis test. Similarly, we construct the modified data
vector first, for simplicity, and our ICS signal is a continuous spectrum, we ignore the line-like backgrounds, the data vector of 
ring i in the energy bin k is
\begin{equation}
y_i^k = d_i^k - A_{sig}\mu_{\rm sig,i}^k - \langle d_i^k - A_{sig}\mu_{\rm sig,i}^k \rangle_i
\label{Eq:data_vector}
\end{equation}
where $\langle \dots \rangle_i$ is averaging over the 8 signal rings, $d_i^k$ is the signal data of ring i in the energy bin k.
Following the work ~\cite{Foster:2021ngm}, $y_i^k$ is described by Gaussian Process (GP) with the kernel:
\begin{equation}
K(E, E') = A_{GP}\exp\left[ -\frac{(E-E')^2}{2EE'\sigma_{E}^2} \right]
\label{Eq:gp_kernel}
\end{equation}
$A_{GP}$ and $\sigma_{E}$ are the hyper parameters, we fix $\sigma_{E} = 0.3$ and free the $A_{GP}$. The GP process marginal likelihood
is given as:
\begin{equation}
\begin{aligned}
\log p(d_k | {\bm \theta}) &= - {1 \over 2} {{\bf y}^k}^T \left[ {\bf K} + (\sigma^k)^2 {\bf I}\right]^{-1} {\bf y}^k\\
&- {1 \over 2} \log | {\bf K} + (\sigma^k)^2 {\bf I} | - {n \over 2} \log(2 \pi) \,.
\label{Eq:likelihood}
\end{aligned}
\end{equation}
the parameters $\bm \theta$ contains the fraction factor $f_{\rm PBH}$ and the nuisance parameters $A_{GP}$, the eye of the diagonal matrix $(\sigma^k)^2 {\bf I}$ is the uncertainty of signal of ring k in energy bins, and $n$ is the number of energy channels.
The test statistic (TS) to determine the best-fit model under a maximum likelihood estimation is given as:
\begin{equation}
\rm TS = -2\left[\log p(d_k | {f_{\rm PBH}, \hat{A}_{GP}}) - \log p_{max}\right]
\label{Eq:TS}
\end{equation}
The first term in the bracket is the maximum likelihood under fraction value $f_{\rm PBH}$, and the second term is the maximum 
likelihood. We search for the $f_{\rm PBH}$ at $95\%$ confidence level, which corresponds to $\rm TS = 2.71$.
Since only PBH with a mass around $1\times 10^{15}$ g produces a massive ICS signal, the constrain of us by using XMM-Newton will only 
concentrate on mass range $[1\times 10^{15}, 5\times 10^{15}]$, we also compare the result with constrain from COMPTEL. We also calculate
the constraint that comes from each ring, where we consider the data vector in Eq.~\ref{Eq:data_vector} independently. We show the results in Fig.~\ref{Fig:ring_result}, and we can find the most stringent bound on PBH fraction is from ring 3. The limits of COMPTEL are quite strong in the mass region $M_{\rm PHB}\geq 3\times 10^{15}$ g, and even stronger than our results, but lose the sensitivity in the mass region $M_{\rm PHB}\leq 3\times 10^{15}$ g. While our analysis set a strong bound on the PBH fraction with $f_{\rm PBH} \leq 10^{-6}$ when $M_{\rm PHB} = 1\times 10^{15}$ g. For a lighter PBH with mass $M_{\rm PHB} < 10^{15}$ g, it is unable to be a dark matter candidate, for the high temperature $T_{\rm BH}$, such a light PBH will soon decay, we ignore this mass region.

{It is evident that both CMB and EGB exhibit superior sensitivity compared to soft X-rays. However, it is important to note that CMB sensitivity is often dependent on assumptions about the early universe and PBH model construction. In contrast, our approach refrains from making any assumptions regarding the early universe, thereby offering an independent means of probing primordial black holes in the late universe. While our method lags behind EGB in sensitivity, it is worth acknowledging the temporal limitations imposed by the XMM satellite, a concern that is anticipated to be ameliorated with future satellite improvements.}

\section{Conclusion}
In this work, we investigate the potential contribution of light primordial black holes (PBHs) to the density of dark matter. Utilizing archival data from the XMM instrument, we analyze soft X-ray emissions resulting from PBH evaporation through inverse Compton scattering in the PBHs mass range from $1\times 10^{15}$ g to $1\times 10^{16}$ g. We find that the advent of new XMM telescopes offers opportunities in the search of PBHs, {which effectively provides an alternative probe for the light PBHs.}

\section*{acknowledgments}
This work was supported by the National Natural Science Foundation of China under Grant Nos. 12275232, 12275134, 12305111 and 12005180, the Natural Science Foundation of Shandong Province under Grant No. ZR2020QA083, the Project of Shandong Province Higher Educational Science and Technology Program under Grant No. 2019KJJ007, and the Project for Young Scientists in Basic Research of Chinese Academy of Sciences under Grant No. YSBR-061.
 
\vspace{-.3cm}
\bibliography{main}

\begin{thebibliography}{44}%
\makeatletter
\providecommand \@ifxundefined [1]{%
 \@ifx{#1\undefined}
}%
\providecommand \@ifnum [1]{%
 \ifnum #1\expandafter \@firstoftwo
 \else \expandafter \@secondoftwo
 \fi
}%
\providecommand \@ifx [1]{%
 \ifx #1\expandafter \@firstoftwo
 \else \expandafter \@secondoftwo
 \fi
}%
\providecommand \natexlab [1]{#1}%
\providecommand \enquote  [1]{``#1''}%
\providecommand \bibnamefont  [1]{#1}%
\providecommand \bibfnamefont [1]{#1}%
\providecommand \citenamefont [1]{#1}%
\providecommand \href@noop [0]{\@secondoftwo}%
\providecommand \href [0]{\begingroup \@sanitize@url \@href}%
\providecommand \@href[1]{\@@startlink{#1}\@@href}%
\providecommand \@@href[1]{\endgroup#1\@@endlink}%
\providecommand \@sanitize@url [0]{\catcode `\\12\catcode `\$12\catcode
  `\&12\catcode `\#12\catcode `\^12\catcode `\_12\catcode `\%12\relax}%
\providecommand \@@startlink[1]{}%
\providecommand \@@endlink[0]{}%
\providecommand \url  [0]{\begingroup\@sanitize@url \@url }%
\providecommand \@url [1]{\endgroup\@href {#1}{\urlprefix }}%
\providecommand \urlprefix  [0]{URL }%
\providecommand \Eprint [0]{\href }%
\providecommand \doibase [0]{http://dx.doi.org/}%
\providecommand \selectlanguage [0]{\@gobble}%
\providecommand \bibinfo  [0]{\@secondoftwo}%
\providecommand \bibfield  [0]{\@secondoftwo}%
\providecommand \translation [1]{[#1]}%
\providecommand \BibitemOpen [0]{}%
\providecommand \bibitemStop [0]{}%
\providecommand \bibitemNoStop [0]{.\EOS\space}%
\providecommand \EOS [0]{\spacefactor3000\relax}%
\providecommand \BibitemShut  [1]{\csname bibitem#1\endcsname}%
\let\auto@bib@innerbib\@empty
%</preamble>
\bibitem [{\citenamefont {Bertone}\ \emph {et~al.}(2005)\citenamefont
  {Bertone}, \citenamefont {Hooper},\ and\ \citenamefont
  {Silk}}]{Bertone:2004pz}%
  \BibitemOpen
  \bibfield  {author} {\bibinfo {author} {\bibfnamefont {G.}~\bibnamefont
  {Bertone}}, \bibinfo {author} {\bibfnamefont {D.}~\bibnamefont {Hooper}}, \
  and\ \bibinfo {author} {\bibfnamefont {J.}~\bibnamefont {Silk}},\ }\href
  {\doibase 10.1016/j.physrep.2004.08.031} {\bibfield  {journal} {\bibinfo
  {journal} {Phys. Rept.}\ }\textbf {\bibinfo {volume} {405}},\ \bibinfo
  {pages} {279} (\bibinfo {year} {2005})},\ \Eprint
  {http://arxiv.org/abs/hep-ph/0404175} {arXiv:hep-ph/0404175} \BibitemShut
  {NoStop}%
\bibitem [{\citenamefont {Zel'dovich}\ and\ \citenamefont
  {Novikov}(1967)}]{Zeldovich:1967lct}%
  \BibitemOpen
  \bibfield  {author} {\bibinfo {author} {\bibfnamefont {Y.~B.}\ \bibnamefont
  {Zel'dovich}}\ and\ \bibinfo {author} {\bibfnamefont {I.~D.}\ \bibnamefont
  {Novikov}},\ }\href@noop {} {\bibfield  {journal} {\bibinfo  {journal}
  {Soviet Astron. AJ (Engl. Transl. ),}\ }\textbf {\bibinfo {volume} {10}},\
  \bibinfo {pages} {602} (\bibinfo {year} {1967})}\BibitemShut {NoStop}%
\bibitem [{\citenamefont {Chapline}(1975)}]{Chapline:1975ojl}%
  \BibitemOpen
  \bibfield  {author} {\bibinfo {author} {\bibfnamefont {G.~F.}\ \bibnamefont
  {Chapline}},\ }\href {\doibase 10.1038/253251a0} {\bibfield  {journal}
  {\bibinfo  {journal} {Nature}\ }\textbf {\bibinfo {volume} {253}},\ \bibinfo
  {pages} {251} (\bibinfo {year} {1975})}\BibitemShut {NoStop}%
\bibitem [{\citenamefont {Carr}\ and\ \citenamefont
  {Lidsey}(1993)}]{Carr:1993aq}%
  \BibitemOpen
  \bibfield  {author} {\bibinfo {author} {\bibfnamefont {B.~J.}\ \bibnamefont
  {Carr}}\ and\ \bibinfo {author} {\bibfnamefont {J.~E.}\ \bibnamefont
  {Lidsey}},\ }\href {\doibase 10.1103/PhysRevD.48.543} {\bibfield  {journal}
  {\bibinfo  {journal} {Phys. Rev. D}\ }\textbf {\bibinfo {volume} {48}},\
  \bibinfo {pages} {543} (\bibinfo {year} {1993})}\BibitemShut {NoStop}%
\bibitem [{\citenamefont {Ivanov}\ \emph {et~al.}(1994)\citenamefont {Ivanov},
  \citenamefont {Naselsky},\ and\ \citenamefont {Novikov}}]{Ivanov:1994pa}%
  \BibitemOpen
  \bibfield  {author} {\bibinfo {author} {\bibfnamefont {P.}~\bibnamefont
  {Ivanov}}, \bibinfo {author} {\bibfnamefont {P.}~\bibnamefont {Naselsky}}, \
  and\ \bibinfo {author} {\bibfnamefont {I.}~\bibnamefont {Novikov}},\ }\href
  {\doibase 10.1103/PhysRevD.50.7173} {\bibfield  {journal} {\bibinfo
  {journal} {Phys. Rev. D}\ }\textbf {\bibinfo {volume} {50}},\ \bibinfo
  {pages} {7173} (\bibinfo {year} {1994})}\BibitemShut {NoStop}%
\bibitem [{\citenamefont {Garcia-Bellido}\ \emph {et~al.}(1996)\citenamefont
  {Garcia-Bellido}, \citenamefont {Linde},\ and\ \citenamefont
  {Wands}}]{Garcia-Bellido:1996mdl}%
  \BibitemOpen
  \bibfield  {author} {\bibinfo {author} {\bibfnamefont {J.}~\bibnamefont
  {Garcia-Bellido}}, \bibinfo {author} {\bibfnamefont {A.~D.}\ \bibnamefont
  {Linde}}, \ and\ \bibinfo {author} {\bibfnamefont {D.}~\bibnamefont
  {Wands}},\ }\href {\doibase 10.1103/PhysRevD.54.6040} {\bibfield  {journal}
  {\bibinfo  {journal} {Phys. Rev. D}\ }\textbf {\bibinfo {volume} {54}},\
  \bibinfo {pages} {6040} (\bibinfo {year} {1996})},\ \Eprint
  {http://arxiv.org/abs/astro-ph/9605094} {arXiv:astro-ph/9605094} \BibitemShut
  {NoStop}%
\bibitem [{\citenamefont {Randall}\ \emph {et~al.}(1996)\citenamefont
  {Randall}, \citenamefont {Soljacic},\ and\ \citenamefont
  {Guth}}]{Randall:1995dj}%
  \BibitemOpen
  \bibfield  {author} {\bibinfo {author} {\bibfnamefont {L.}~\bibnamefont
  {Randall}}, \bibinfo {author} {\bibfnamefont {M.}~\bibnamefont {Soljacic}}, \
  and\ \bibinfo {author} {\bibfnamefont {A.~H.}\ \bibnamefont {Guth}},\ }\href
  {\doibase 10.1016/0550-3213(96)00174-5} {\bibfield  {journal} {\bibinfo
  {journal} {Nucl. Phys. B}\ }\textbf {\bibinfo {volume} {472}},\ \bibinfo
  {pages} {377} (\bibinfo {year} {1996})},\ \Eprint
  {http://arxiv.org/abs/hep-ph/9512439} {arXiv:hep-ph/9512439} \BibitemShut
  {NoStop}%
\bibitem [{\citenamefont {Niikura}\ \emph
  {et~al.}(2019{\natexlab{a}})\citenamefont {Niikura}, \citenamefont {Takada},
  \citenamefont {Yokoyama}, \citenamefont {Sumi},\ and\ \citenamefont
  {Masaki}}]{Niikura:2019kqi}%
  \BibitemOpen
  \bibfield  {author} {\bibinfo {author} {\bibfnamefont {H.}~\bibnamefont
  {Niikura}}, \bibinfo {author} {\bibfnamefont {M.}~\bibnamefont {Takada}},
  \bibinfo {author} {\bibfnamefont {S.}~\bibnamefont {Yokoyama}}, \bibinfo
  {author} {\bibfnamefont {T.}~\bibnamefont {Sumi}}, \ and\ \bibinfo {author}
  {\bibfnamefont {S.}~\bibnamefont {Masaki}},\ }\href {\doibase
  10.1103/PhysRevD.99.083503} {\bibfield  {journal} {\bibinfo  {journal} {Phys.
  Rev. D}\ }\textbf {\bibinfo {volume} {99}},\ \bibinfo {pages} {083503}
  (\bibinfo {year} {2019}{\natexlab{a}})},\ \Eprint
  {http://arxiv.org/abs/1901.07120} {arXiv:1901.07120 [astro-ph.CO]}
  \BibitemShut {NoStop}%
\bibitem [{\citenamefont {Mediavilla}\ \emph {et~al.}(2017)\citenamefont
  {Mediavilla}, \citenamefont {Jim\'enez-Vicente}, \citenamefont {Mu\~noz},
  \citenamefont {Vives-Arias},\ and\ \citenamefont
  {Calder\'on-Infante}}]{Mediavilla:2017bok}%
  \BibitemOpen
  \bibfield  {author} {\bibinfo {author} {\bibfnamefont {E.}~\bibnamefont
  {Mediavilla}}, \bibinfo {author} {\bibfnamefont {J.}~\bibnamefont
  {Jim\'enez-Vicente}}, \bibinfo {author} {\bibfnamefont {J.~A.}\ \bibnamefont
  {Mu\~noz}}, \bibinfo {author} {\bibfnamefont {H.}~\bibnamefont
  {Vives-Arias}}, \ and\ \bibinfo {author} {\bibfnamefont {J.}~\bibnamefont
  {Calder\'on-Infante}},\ }\href {\doibase 10.3847/2041-8213/aa5dab} {\bibfield
   {journal} {\bibinfo  {journal} {Astrophys. J. Lett.}\ }\textbf {\bibinfo
  {volume} {836}},\ \bibinfo {pages} {L18} (\bibinfo {year} {2017})},\ \Eprint
  {http://arxiv.org/abs/1702.00947} {arXiv:1702.00947 [astro-ph.GA]}
  \BibitemShut {NoStop}%
\bibitem [{\citenamefont {Allsman}\ \emph {et~al.}(2001)\citenamefont {Allsman}
  \emph {et~al.}}]{Macho:2000nvd}%
  \BibitemOpen
  \bibfield  {author} {\bibinfo {author} {\bibfnamefont {R.~A.}\ \bibnamefont
  {Allsman}} \emph {et~al.} (\bibinfo {collaboration} {Macho}),\ }\href
  {\doibase 10.1086/319636} {\bibfield  {journal} {\bibinfo  {journal}
  {Astrophys. J. Lett.}\ }\textbf {\bibinfo {volume} {550}},\ \bibinfo {pages}
  {L169} (\bibinfo {year} {2001})},\ \Eprint
  {http://arxiv.org/abs/astro-ph/0011506} {arXiv:astro-ph/0011506} \BibitemShut
  {NoStop}%
\bibitem [{\citenamefont {Griest}\ \emph {et~al.}(2014)\citenamefont {Griest},
  \citenamefont {Cieplak},\ and\ \citenamefont {Lehner}}]{Griest:2013aaa}%
  \BibitemOpen
  \bibfield  {author} {\bibinfo {author} {\bibfnamefont {K.}~\bibnamefont
  {Griest}}, \bibinfo {author} {\bibfnamefont {A.~M.}\ \bibnamefont {Cieplak}},
  \ and\ \bibinfo {author} {\bibfnamefont {M.~J.}\ \bibnamefont {Lehner}},\
  }\href {\doibase 10.1088/0004-637X/786/2/158} {\bibfield  {journal} {\bibinfo
   {journal} {Astrophys. J.}\ }\textbf {\bibinfo {volume} {786}},\ \bibinfo
  {pages} {158} (\bibinfo {year} {2014})},\ \Eprint
  {http://arxiv.org/abs/1307.5798} {arXiv:1307.5798 [astro-ph.CO]} \BibitemShut
  {NoStop}%
\bibitem [{\citenamefont {Niikura}\ \emph
  {et~al.}(2019{\natexlab{b}})\citenamefont {Niikura} \emph
  {et~al.}}]{Niikura:2017zjd}%
  \BibitemOpen
  \bibfield  {author} {\bibinfo {author} {\bibfnamefont {H.}~\bibnamefont
  {Niikura}} \emph {et~al.},\ }\href {\doibase 10.1038/s41550-019-0723-1}
  {\bibfield  {journal} {\bibinfo  {journal} {Nature Astron.}\ }\textbf
  {\bibinfo {volume} {3}},\ \bibinfo {pages} {524} (\bibinfo {year}
  {2019}{\natexlab{b}})},\ \Eprint {http://arxiv.org/abs/1701.02151}
  {arXiv:1701.02151 [astro-ph.CO]} \BibitemShut {NoStop}%
\bibitem [{\citenamefont {Peta\v{c}}\ \emph {et~al.}(2022)\citenamefont
  {Peta\v{c}}, \citenamefont {Lavalle},\ and\ \citenamefont
  {Jedamzik}}]{Petac:2022rio}%
  \BibitemOpen
  \bibfield  {author} {\bibinfo {author} {\bibfnamefont {M.}~\bibnamefont
  {Peta\v{c}}}, \bibinfo {author} {\bibfnamefont {J.}~\bibnamefont {Lavalle}},
  \ and\ \bibinfo {author} {\bibfnamefont {K.}~\bibnamefont {Jedamzik}},\
  }\href {\doibase 10.1103/PhysRevD.105.083520} {\bibfield  {journal} {\bibinfo
   {journal} {Phys. Rev. D}\ }\textbf {\bibinfo {volume} {105}},\ \bibinfo
  {pages} {083520} (\bibinfo {year} {2022})},\ \Eprint
  {http://arxiv.org/abs/2201.02521} {arXiv:2201.02521 [astro-ph.CO]}
  \BibitemShut {NoStop}%
\bibitem [{\citenamefont {Ali-Ha\"\i{}moud}\ and\ \citenamefont
  {Kamionkowski}(2017)}]{Ali-Haimoud:2016mbv}%
  \BibitemOpen
  \bibfield  {author} {\bibinfo {author} {\bibfnamefont {Y.}~\bibnamefont
  {Ali-Ha\"\i{}moud}}\ and\ \bibinfo {author} {\bibfnamefont {M.}~\bibnamefont
  {Kamionkowski}},\ }\href {\doibase 10.1103/PhysRevD.95.043534} {\bibfield
  {journal} {\bibinfo  {journal} {Phys. Rev. D}\ }\textbf {\bibinfo {volume}
  {95}},\ \bibinfo {pages} {043534} (\bibinfo {year} {2017})},\ \Eprint
  {http://arxiv.org/abs/1612.05644} {arXiv:1612.05644 [astro-ph.CO]}
  \BibitemShut {NoStop}%
\bibitem [{\citenamefont {Poulter}\ \emph {et~al.}(2019)\citenamefont
  {Poulter}, \citenamefont {Ali-Ha\"\i{}moud}, \citenamefont {Hamann},
  \citenamefont {White},\ and\ \citenamefont {Williams}}]{Poulter:2019ooo}%
  \BibitemOpen
  \bibfield  {author} {\bibinfo {author} {\bibfnamefont {H.}~\bibnamefont
  {Poulter}}, \bibinfo {author} {\bibfnamefont {Y.}~\bibnamefont
  {Ali-Ha\"\i{}moud}}, \bibinfo {author} {\bibfnamefont {J.}~\bibnamefont
  {Hamann}}, \bibinfo {author} {\bibfnamefont {M.}~\bibnamefont {White}}, \
  and\ \bibinfo {author} {\bibfnamefont {A.~G.}\ \bibnamefont {Williams}},\
  }\href@noop {} {\  (\bibinfo {year} {2019})},\ \Eprint
  {http://arxiv.org/abs/1907.06485} {arXiv:1907.06485 [astro-ph.CO]}
  \BibitemShut {NoStop}%
\bibitem [{\citenamefont {Carr}\ \emph {et~al.}(2010)\citenamefont {Carr},
  \citenamefont {Kohri}, \citenamefont {Sendouda},\ and\ \citenamefont
  {Yokoyama}}]{Carr:2009jm}%
  \BibitemOpen
  \bibfield  {author} {\bibinfo {author} {\bibfnamefont {B.~J.}\ \bibnamefont
  {Carr}}, \bibinfo {author} {\bibfnamefont {K.}~\bibnamefont {Kohri}},
  \bibinfo {author} {\bibfnamefont {Y.}~\bibnamefont {Sendouda}}, \ and\
  \bibinfo {author} {\bibfnamefont {J.}~\bibnamefont {Yokoyama}},\ }\href
  {\doibase 10.1103/PhysRevD.81.104019} {\bibfield  {journal} {\bibinfo
  {journal} {Phys. Rev. D}\ }\textbf {\bibinfo {volume} {81}},\ \bibinfo
  {pages} {104019} (\bibinfo {year} {2010})},\ \Eprint
  {http://arxiv.org/abs/0912.5297} {arXiv:0912.5297 [astro-ph.CO]} \BibitemShut
  {NoStop}%
\bibitem [{\citenamefont {Lehmann}\ \emph {et~al.}(2018)\citenamefont
  {Lehmann}, \citenamefont {Profumo},\ and\ \citenamefont
  {Yant}}]{Lehmann:2018ejc}%
  \BibitemOpen
  \bibfield  {author} {\bibinfo {author} {\bibfnamefont {B.~V.}\ \bibnamefont
  {Lehmann}}, \bibinfo {author} {\bibfnamefont {S.}~\bibnamefont {Profumo}}, \
  and\ \bibinfo {author} {\bibfnamefont {J.}~\bibnamefont {Yant}},\ }\href
  {\doibase 10.1088/1475-7516/2018/04/007} {\bibfield  {journal} {\bibinfo
  {journal} {JCAP}\ }\textbf {\bibinfo {volume} {04}},\ \bibinfo {pages} {007}
  (\bibinfo {year} {2018})},\ \Eprint {http://arxiv.org/abs/1801.00808}
  {arXiv:1801.00808 [astro-ph.CO]} \BibitemShut {NoStop}%
\bibitem [{\citenamefont {Coogan}\ \emph {et~al.}(2021)\citenamefont {Coogan},
  \citenamefont {Morrison},\ and\ \citenamefont {Profumo}}]{Coogan:2020tuf}%
  \BibitemOpen
  \bibfield  {author} {\bibinfo {author} {\bibfnamefont {A.}~\bibnamefont
  {Coogan}}, \bibinfo {author} {\bibfnamefont {L.}~\bibnamefont {Morrison}}, \
  and\ \bibinfo {author} {\bibfnamefont {S.}~\bibnamefont {Profumo}},\ }\href
  {\doibase 10.1103/PhysRevLett.126.171101} {\bibfield  {journal} {\bibinfo
  {journal} {Phys. Rev. Lett.}\ }\textbf {\bibinfo {volume} {126}},\ \bibinfo
  {pages} {171101} (\bibinfo {year} {2021})},\ \Eprint
  {http://arxiv.org/abs/2010.04797} {arXiv:2010.04797 [astro-ph.CO]}
  \BibitemShut {NoStop}%
\bibitem [{\citenamefont {Capanema}\ \emph {et~al.}(2021)\citenamefont
  {Capanema}, \citenamefont {Esmaeili},\ and\ \citenamefont
  {Esmaili}}]{Capanema:2021hnm}%
  \BibitemOpen
  \bibfield  {author} {\bibinfo {author} {\bibfnamefont {A.}~\bibnamefont
  {Capanema}}, \bibinfo {author} {\bibfnamefont {A.}~\bibnamefont {Esmaeili}},
  \ and\ \bibinfo {author} {\bibfnamefont {A.}~\bibnamefont {Esmaili}},\ }\href
  {\doibase 10.1088/1475-7516/2021/12/051} {\bibfield  {journal} {\bibinfo
  {journal} {JCAP}\ }\textbf {\bibinfo {volume} {12}},\ \bibinfo {pages} {051}
  (\bibinfo {year} {2021})},\ \Eprint {http://arxiv.org/abs/2110.05637}
  {arXiv:2110.05637 [hep-ph]} \BibitemShut {NoStop}%
\bibitem [{\citenamefont {Chen}\ \emph {et~al.}(2022)\citenamefont {Chen},
  \citenamefont {Zhang},\ and\ \citenamefont {Long}}]{Chen:2021ngo}%
  \BibitemOpen
  \bibfield  {author} {\bibinfo {author} {\bibfnamefont {S.}~\bibnamefont
  {Chen}}, \bibinfo {author} {\bibfnamefont {H.-H.}\ \bibnamefont {Zhang}}, \
  and\ \bibinfo {author} {\bibfnamefont {G.}~\bibnamefont {Long}},\ }\href
  {\doibase 10.1103/PhysRevD.105.063008} {\bibfield  {journal} {\bibinfo
  {journal} {Phys. Rev. D}\ }\textbf {\bibinfo {volume} {105}},\ \bibinfo
  {pages} {063008} (\bibinfo {year} {2022})},\ \Eprint
  {http://arxiv.org/abs/2112.15463} {arXiv:2112.15463 [astro-ph.CO]}
  \BibitemShut {NoStop}%
\bibitem [{\citenamefont {Agashe}\ \emph {et~al.}(2022)\citenamefont {Agashe},
  \citenamefont {Chang}, \citenamefont {Clark}, \citenamefont {Dutta},
  \citenamefont {Tsai},\ and\ \citenamefont {Xu}}]{Agashe:2022jgk}%
  \BibitemOpen
  \bibfield  {author} {\bibinfo {author} {\bibfnamefont {K.}~\bibnamefont
  {Agashe}}, \bibinfo {author} {\bibfnamefont {J.~H.}\ \bibnamefont {Chang}},
  \bibinfo {author} {\bibfnamefont {S.~J.}\ \bibnamefont {Clark}}, \bibinfo
  {author} {\bibfnamefont {B.}~\bibnamefont {Dutta}}, \bibinfo {author}
  {\bibfnamefont {Y.}~\bibnamefont {Tsai}}, \ and\ \bibinfo {author}
  {\bibfnamefont {T.}~\bibnamefont {Xu}},\ }\href {\doibase
  10.1103/PhysRevD.105.123009} {\bibfield  {journal} {\bibinfo  {journal}
  {Phys. Rev. D}\ }\textbf {\bibinfo {volume} {105}},\ \bibinfo {pages}
  {123009} (\bibinfo {year} {2022})},\ \Eprint
  {http://arxiv.org/abs/2202.04653} {arXiv:2202.04653 [astro-ph.CO]}
  \BibitemShut {NoStop}%
\bibitem [{\citenamefont {Xie}(2023)}]{Xie:2023cwi}%
  \BibitemOpen
  \bibfield  {author} {\bibinfo {author} {\bibfnamefont {K.-P.}\ \bibnamefont
  {Xie}},\ }\href {\doibase 10.1088/1475-7516/2023/06/008} {\bibfield
  {journal} {\bibinfo  {journal} {JCAP}\ }\textbf {\bibinfo {volume} {06}},\
  \bibinfo {pages} {008} (\bibinfo {year} {2023})},\ \Eprint
  {http://arxiv.org/abs/2301.02352} {arXiv:2301.02352 [astro-ph.CO]}
  \BibitemShut {NoStop}%
\bibitem [{\citenamefont {Korwar}\ and\ \citenamefont
  {Profumo}(2023)}]{Korwar:2023kpy}%
  \BibitemOpen
  \bibfield  {author} {\bibinfo {author} {\bibfnamefont {M.}~\bibnamefont
  {Korwar}}\ and\ \bibinfo {author} {\bibfnamefont {S.}~\bibnamefont
  {Profumo}},\ }\href {\doibase 10.1088/1475-7516/2023/05/054} {\bibfield
  {journal} {\bibinfo  {journal} {JCAP}\ }\textbf {\bibinfo {volume} {05}},\
  \bibinfo {pages} {054} (\bibinfo {year} {2023})},\ \Eprint
  {http://arxiv.org/abs/2302.04408} {arXiv:2302.04408 [hep-ph]} \BibitemShut
  {NoStop}%
\bibitem [{\citenamefont {Jung}\ and\ \citenamefont
  {Shin}(2019)}]{Jung:2017flg}%
  \BibitemOpen
  \bibfield  {author} {\bibinfo {author} {\bibfnamefont {S.}~\bibnamefont
  {Jung}}\ and\ \bibinfo {author} {\bibfnamefont {C.~S.}\ \bibnamefont
  {Shin}},\ }\href {\doibase 10.1103/PhysRevLett.122.041103} {\bibfield
  {journal} {\bibinfo  {journal} {Phys. Rev. Lett.}\ }\textbf {\bibinfo
  {volume} {122}},\ \bibinfo {pages} {041103} (\bibinfo {year} {2019})},\
  \Eprint {http://arxiv.org/abs/1712.01396} {arXiv:1712.01396 [astro-ph.CO]}
  \BibitemShut {NoStop}%
\bibitem [{\citenamefont {Chen}\ and\ \citenamefont
  {Huang}(2020)}]{Chen:2019irf}%
  \BibitemOpen
  \bibfield  {author} {\bibinfo {author} {\bibfnamefont {Z.-C.}\ \bibnamefont
  {Chen}}\ and\ \bibinfo {author} {\bibfnamefont {Q.-G.}\ \bibnamefont
  {Huang}},\ }\href {\doibase 10.1088/1475-7516/2020/08/039} {\bibfield
  {journal} {\bibinfo  {journal} {JCAP}\ }\textbf {\bibinfo {volume} {08}},\
  \bibinfo {pages} {039} (\bibinfo {year} {2020})},\ \Eprint
  {http://arxiv.org/abs/1904.02396} {arXiv:1904.02396 [astro-ph.CO]}
  \BibitemShut {NoStop}%
\bibitem [{\citenamefont {Abbott}\ \emph {et~al.}(2019)\citenamefont {Abbott}
  \emph {et~al.}}]{LIGOScientific:2019kan}%
  \BibitemOpen
  \bibfield  {author} {\bibinfo {author} {\bibfnamefont {B.~P.}\ \bibnamefont
  {Abbott}} \emph {et~al.} (\bibinfo {collaboration} {LIGO Scientific,
  Virgo}),\ }\href {\doibase 10.1103/PhysRevLett.123.161102} {\bibfield
  {journal} {\bibinfo  {journal} {Phys. Rev. Lett.}\ }\textbf {\bibinfo
  {volume} {123}},\ \bibinfo {pages} {161102} (\bibinfo {year} {2019})},\
  \Eprint {http://arxiv.org/abs/1904.08976} {arXiv:1904.08976 [astro-ph.CO]}
  \BibitemShut {NoStop}%
\bibitem [{\citenamefont {Kavanagh}\ \emph {et~al.}(2018)\citenamefont
  {Kavanagh}, \citenamefont {Gaggero},\ and\ \citenamefont
  {Bertone}}]{Kavanagh:2018ggo}%
  \BibitemOpen
  \bibfield  {author} {\bibinfo {author} {\bibfnamefont {B.~J.}\ \bibnamefont
  {Kavanagh}}, \bibinfo {author} {\bibfnamefont {D.}~\bibnamefont {Gaggero}}, \
  and\ \bibinfo {author} {\bibfnamefont {G.}~\bibnamefont {Bertone}},\ }\href
  {\doibase 10.1103/PhysRevD.98.023536} {\bibfield  {journal} {\bibinfo
  {journal} {Phys. Rev. D}\ }\textbf {\bibinfo {volume} {98}},\ \bibinfo
  {pages} {023536} (\bibinfo {year} {2018})},\ \Eprint
  {http://arxiv.org/abs/1805.09034} {arXiv:1805.09034 [astro-ph.CO]}
  \BibitemShut {NoStop}%
\bibitem [{\citenamefont {Chen}\ \emph {et~al.}(2020)\citenamefont {Chen},
  \citenamefont {Yuan},\ and\ \citenamefont {Huang}}]{Chen:2019xse}%
  \BibitemOpen
  \bibfield  {author} {\bibinfo {author} {\bibfnamefont {Z.-C.}\ \bibnamefont
  {Chen}}, \bibinfo {author} {\bibfnamefont {C.}~\bibnamefont {Yuan}}, \ and\
  \bibinfo {author} {\bibfnamefont {Q.-G.}\ \bibnamefont {Huang}},\ }\href
  {\doibase 10.1103/PhysRevLett.124.251101} {\bibfield  {journal} {\bibinfo
  {journal} {Phys. Rev. Lett.}\ }\textbf {\bibinfo {volume} {124}},\ \bibinfo
  {pages} {251101} (\bibinfo {year} {2020})},\ \Eprint
  {http://arxiv.org/abs/1910.12239} {arXiv:1910.12239 [astro-ph.CO]}
  \BibitemShut {NoStop}%
\bibitem [{\citenamefont {Wang}\ \emph {et~al.}(2020)\citenamefont {Wang},
  \citenamefont {Huang}, \citenamefont {Li},\ and\ \citenamefont
  {Liao}}]{Wang:2019kzb}%
  \BibitemOpen
  \bibfield  {author} {\bibinfo {author} {\bibfnamefont {Y.-F.}\ \bibnamefont
  {Wang}}, \bibinfo {author} {\bibfnamefont {Q.-G.}\ \bibnamefont {Huang}},
  \bibinfo {author} {\bibfnamefont {T.~G.~F.}\ \bibnamefont {Li}}, \ and\
  \bibinfo {author} {\bibfnamefont {S.}~\bibnamefont {Liao}},\ }\href {\doibase
  10.1103/PhysRevD.101.063019} {\bibfield  {journal} {\bibinfo  {journal}
  {Phys. Rev. D}\ }\textbf {\bibinfo {volume} {101}},\ \bibinfo {pages}
  {063019} (\bibinfo {year} {2020})},\ \Eprint
  {http://arxiv.org/abs/1910.07397} {arXiv:1910.07397 [astro-ph.CO]}
  \BibitemShut {NoStop}%
\bibitem [{\citenamefont {Dror}\ \emph {et~al.}(2019)\citenamefont {Dror},
  \citenamefont {Ramani}, \citenamefont {Trickle},\ and\ \citenamefont
  {Zurek}}]{Dror:2019twh}%
  \BibitemOpen
  \bibfield  {author} {\bibinfo {author} {\bibfnamefont {J.~A.}\ \bibnamefont
  {Dror}}, \bibinfo {author} {\bibfnamefont {H.}~\bibnamefont {Ramani}},
  \bibinfo {author} {\bibfnamefont {T.}~\bibnamefont {Trickle}}, \ and\
  \bibinfo {author} {\bibfnamefont {K.~M.}\ \bibnamefont {Zurek}},\ }\href
  {\doibase 10.1103/PhysRevD.100.023003} {\bibfield  {journal} {\bibinfo
  {journal} {Phys. Rev. D}\ }\textbf {\bibinfo {volume} {100}},\ \bibinfo
  {pages} {023003} (\bibinfo {year} {2019})},\ \Eprint
  {http://arxiv.org/abs/1901.04490} {arXiv:1901.04490 [astro-ph.CO]}
  \BibitemShut {NoStop}%
\bibitem [{\citenamefont {Capela}\ \emph {et~al.}(2013)\citenamefont {Capela},
  \citenamefont {Pshirkov},\ and\ \citenamefont {Tinyakov}}]{Capela:2013yf}%
  \BibitemOpen
  \bibfield  {author} {\bibinfo {author} {\bibfnamefont {F.}~\bibnamefont
  {Capela}}, \bibinfo {author} {\bibfnamefont {M.}~\bibnamefont {Pshirkov}}, \
  and\ \bibinfo {author} {\bibfnamefont {P.}~\bibnamefont {Tinyakov}},\ }\href
  {\doibase 10.1103/PhysRevD.87.123524} {\bibfield  {journal} {\bibinfo
  {journal} {Phys. Rev. D}\ }\textbf {\bibinfo {volume} {87}},\ \bibinfo
  {pages} {123524} (\bibinfo {year} {2013})},\ \Eprint
  {http://arxiv.org/abs/1301.4984} {arXiv:1301.4984 [astro-ph.CO]} \BibitemShut
  {NoStop}%
\bibitem [{\citenamefont {Graham}\ \emph {et~al.}(2015)\citenamefont {Graham},
  \citenamefont {Rajendran},\ and\ \citenamefont {Varela}}]{Graham:2015apa}%
  \BibitemOpen
  \bibfield  {author} {\bibinfo {author} {\bibfnamefont {P.~W.}\ \bibnamefont
  {Graham}}, \bibinfo {author} {\bibfnamefont {S.}~\bibnamefont {Rajendran}}, \
  and\ \bibinfo {author} {\bibfnamefont {J.}~\bibnamefont {Varela}},\ }\href
  {\doibase 10.1103/PhysRevD.92.063007} {\bibfield  {journal} {\bibinfo
  {journal} {Phys. Rev. D}\ }\textbf {\bibinfo {volume} {92}},\ \bibinfo
  {pages} {063007} (\bibinfo {year} {2015})},\ \Eprint
  {http://arxiv.org/abs/1505.04444} {arXiv:1505.04444 [hep-ph]} \BibitemShut
  {NoStop}%
\bibitem [{\citenamefont {Lu}\ and\ \citenamefont {Wu}(2019)}]{Lu:2019ktw}%
  \BibitemOpen
  \bibfield  {author} {\bibinfo {author} {\bibfnamefont {B.-Q.}\ \bibnamefont
  {Lu}}\ and\ \bibinfo {author} {\bibfnamefont {Y.-L.}\ \bibnamefont {Wu}},\
  }\href {\doibase 10.1103/PhysRevD.99.123023} {\bibfield  {journal} {\bibinfo
  {journal} {Phys. Rev. D}\ }\textbf {\bibinfo {volume} {99}},\ \bibinfo
  {pages} {123023} (\bibinfo {year} {2019})},\ \Eprint
  {http://arxiv.org/abs/1906.10463} {arXiv:1906.10463 [astro-ph.HE]}
  \BibitemShut {NoStop}%
\bibitem [{\citenamefont {Kierans}(2020)}]{Kierans:2020otl}%
  \BibitemOpen
  \bibfield  {author} {\bibinfo {author} {\bibfnamefont {C.~A.}\ \bibnamefont
  {Kierans}} (\bibinfo {collaboration} {AMEGO Team}),\ }\href {\doibase
  10.1117/12.2562352} {\bibfield  {journal} {\bibinfo  {journal} {Proc. SPIE
  Int. Soc. Opt. Eng.}\ }\textbf {\bibinfo {volume} {11444}},\ \bibinfo {pages}
  {1144431} (\bibinfo {year} {2020})},\ \Eprint
  {http://arxiv.org/abs/2101.03105} {arXiv:2101.03105 [astro-ph.IM]}
  \BibitemShut {NoStop}%
\bibitem [{\citenamefont {Tavani}\ \emph {et~al.}(2018)\citenamefont {Tavani}
  \emph {et~al.}}]{e-ASTROGAM:2017pxr}%
  \BibitemOpen
  \bibfield  {author} {\bibinfo {author} {\bibfnamefont {M.}~\bibnamefont
  {Tavani}} \emph {et~al.} (\bibinfo {collaboration} {e-ASTROGAM}),\ }\href
  {\doibase 10.1016/j.jheap.2018.07.001} {\bibfield  {journal} {\bibinfo
  {journal} {JHEAp}\ }\textbf {\bibinfo {volume} {19}},\ \bibinfo {pages} {1}
  (\bibinfo {year} {2018})},\ \Eprint {http://arxiv.org/abs/1711.01265}
  {arXiv:1711.01265 [astro-ph.HE]} \BibitemShut {NoStop}%
\bibitem [{\citenamefont {Orlando}\ \emph {et~al.}(2022)\citenamefont {Orlando}
  \emph {et~al.}}]{Orlando:2021get}%
  \BibitemOpen
  \bibfield  {author} {\bibinfo {author} {\bibfnamefont {E.}~\bibnamefont
  {Orlando}} \emph {et~al.},\ }\href {\doibase 10.1088/1475-7516/2022/07/036}
  {\bibfield  {journal} {\bibinfo  {journal} {JCAP}\ }\textbf {\bibinfo
  {volume} {07}},\ \bibinfo {pages} {036} (\bibinfo {year} {2022})},\ \Eprint
  {http://arxiv.org/abs/2112.07190} {arXiv:2112.07190 [astro-ph.HE]}
  \BibitemShut {NoStop}%
\bibitem [{\citenamefont {Saha}\ and\ \citenamefont
  {Laha}(2022)}]{Saha:2021pqf}%
  \BibitemOpen
  \bibfield  {author} {\bibinfo {author} {\bibfnamefont {A.~K.}\ \bibnamefont
  {Saha}}\ and\ \bibinfo {author} {\bibfnamefont {R.}~\bibnamefont {Laha}},\
  }\href {\doibase 10.1103/PhysRevD.105.103026} {\bibfield  {journal} {\bibinfo
   {journal} {Phys. Rev. D}\ }\textbf {\bibinfo {volume} {105}},\ \bibinfo
  {pages} {103026} (\bibinfo {year} {2022})},\ \Eprint
  {http://arxiv.org/abs/2112.10794} {arXiv:2112.10794 [astro-ph.CO]}
  \BibitemShut {NoStop}%
\bibitem [{\citenamefont {Bowman}\ \emph {et~al.}(2018)\citenamefont {Bowman},
  \citenamefont {Rogers}, \citenamefont {Monsalve}, \citenamefont {Mozdzen},\
  and\ \citenamefont {Mahesh}}]{Bowman:2018yin}%
  \BibitemOpen
  \bibfield  {author} {\bibinfo {author} {\bibfnamefont {J.~D.}\ \bibnamefont
  {Bowman}}, \bibinfo {author} {\bibfnamefont {A.~E.~E.}\ \bibnamefont
  {Rogers}}, \bibinfo {author} {\bibfnamefont {R.~A.}\ \bibnamefont
  {Monsalve}}, \bibinfo {author} {\bibfnamefont {T.~J.}\ \bibnamefont
  {Mozdzen}}, \ and\ \bibinfo {author} {\bibfnamefont {N.}~\bibnamefont
  {Mahesh}},\ }\href {\doibase 10.1038/nature25792} {\bibfield  {journal}
  {\bibinfo  {journal} {Nature}\ }\textbf {\bibinfo {volume} {555}},\ \bibinfo
  {pages} {67} (\bibinfo {year} {2018})},\ \Eprint
  {http://arxiv.org/abs/1810.05912} {arXiv:1810.05912 [astro-ph.CO]}
  \BibitemShut {NoStop}%
\bibitem [{\citenamefont {Singh}\ \emph {et~al.}(2022)\citenamefont {Singh},
  \citenamefont {Nambissan~T.}, \citenamefont {Subrahmanyan}, \citenamefont
  {Udaya~Shankar}, \citenamefont {Girish}, \citenamefont {Raghunathan},
  \citenamefont {Somashekar}, \citenamefont {Srivani},\ and\ \citenamefont
  {Sathyanarayana~Rao}}]{Singh:2021mxo}%
  \BibitemOpen
  \bibfield  {author} {\bibinfo {author} {\bibfnamefont {S.}~\bibnamefont
  {Singh}}, \bibinfo {author} {\bibfnamefont {J.}~\bibnamefont {Nambissan~T.}},
  \bibinfo {author} {\bibfnamefont {R.}~\bibnamefont {Subrahmanyan}}, \bibinfo
  {author} {\bibfnamefont {N.}~\bibnamefont {Udaya~Shankar}}, \bibinfo {author}
  {\bibfnamefont {B.~S.}\ \bibnamefont {Girish}}, \bibinfo {author}
  {\bibfnamefont {A.}~\bibnamefont {Raghunathan}}, \bibinfo {author}
  {\bibfnamefont {R.}~\bibnamefont {Somashekar}}, \bibinfo {author}
  {\bibfnamefont {K.~S.}\ \bibnamefont {Srivani}}, \ and\ \bibinfo {author}
  {\bibfnamefont {M.}~\bibnamefont {Sathyanarayana~Rao}},\ }\href {\doibase
  10.1038/s41550-022-01610-5} {\bibfield  {journal} {\bibinfo  {journal}
  {Nature Astron.}\ }\textbf {\bibinfo {volume} {6}},\ \bibinfo {pages} {607}
  (\bibinfo {year} {2022})},\ \Eprint {http://arxiv.org/abs/2112.06778}
  {arXiv:2112.06778 [astro-ph.CO]} \BibitemShut {NoStop}%
\bibitem [{\citenamefont {Mukhopadhyay}\ \emph {et~al.}(2021)\citenamefont
  {Mukhopadhyay}, \citenamefont {Majumdar},\ and\ \citenamefont
  {Paul}}]{Mukhopadhyay:2021puu}%
  \BibitemOpen
  \bibfield  {author} {\bibinfo {author} {\bibfnamefont {U.}~\bibnamefont
  {Mukhopadhyay}}, \bibinfo {author} {\bibfnamefont {D.}~\bibnamefont
  {Majumdar}}, \ and\ \bibinfo {author} {\bibfnamefont {A.}~\bibnamefont
  {Paul}},\ }\href@noop {} {\  (\bibinfo {year} {2021})},\ \Eprint
  {http://arxiv.org/abs/2109.14955} {arXiv:2109.14955 [astro-ph.CO]}
  \BibitemShut {NoStop}%
\bibitem [{\citenamefont {Foster}\ \emph {et~al.}(2021)\citenamefont {Foster},
  \citenamefont {Kongsore}, \citenamefont {Dessert}, \citenamefont {Park},
  \citenamefont {Rodd}, \citenamefont {Cranmer},\ and\ \citenamefont
  {Safdi}}]{Foster:2021ngm}%
  \BibitemOpen
  \bibfield  {author} {\bibinfo {author} {\bibfnamefont {J.~W.}\ \bibnamefont
  {Foster}}, \bibinfo {author} {\bibfnamefont {M.}~\bibnamefont {Kongsore}},
  \bibinfo {author} {\bibfnamefont {C.}~\bibnamefont {Dessert}}, \bibinfo
  {author} {\bibfnamefont {Y.}~\bibnamefont {Park}}, \bibinfo {author}
  {\bibfnamefont {N.~L.}\ \bibnamefont {Rodd}}, \bibinfo {author}
  {\bibfnamefont {K.}~\bibnamefont {Cranmer}}, \ and\ \bibinfo {author}
  {\bibfnamefont {B.~R.}\ \bibnamefont {Safdi}},\ }\href {\doibase
  10.1103/PhysRevLett.127.051101} {\bibfield  {journal} {\bibinfo  {journal}
  {Phys. Rev. Lett.}\ }\textbf {\bibinfo {volume} {127}},\ \bibinfo {pages}
  {051101} (\bibinfo {year} {2021})},\ \Eprint
  {http://arxiv.org/abs/2102.02207} {arXiv:2102.02207 [astro-ph.CO]}
  \BibitemShut {NoStop}%
\bibitem [{\citenamefont {Cirelli}\ \emph {et~al.}(2023)\citenamefont
  {Cirelli}, \citenamefont {Fornengo}, \citenamefont {Koechler}, \citenamefont
  {Pinetti},\ and\ \citenamefont {Roach}}]{Cirelli:2023tnx}%
  \BibitemOpen
  \bibfield  {author} {\bibinfo {author} {\bibfnamefont {M.}~\bibnamefont
  {Cirelli}}, \bibinfo {author} {\bibfnamefont {N.}~\bibnamefont {Fornengo}},
  \bibinfo {author} {\bibfnamefont {J.}~\bibnamefont {Koechler}}, \bibinfo
  {author} {\bibfnamefont {E.}~\bibnamefont {Pinetti}}, \ and\ \bibinfo
  {author} {\bibfnamefont {B.~M.}\ \bibnamefont {Roach}},\ }\href {\doibase
  10.1088/1475-7516/2023/07/026} {\bibfield  {journal} {\bibinfo  {journal}
  {JCAP}\ }\textbf {\bibinfo {volume} {07}},\ \bibinfo {pages} {026} (\bibinfo
  {year} {2023})},\ \Eprint {http://arxiv.org/abs/2303.08854} {arXiv:2303.08854
  [hep-ph]} \BibitemShut {NoStop}%
\bibitem [{\citenamefont {De~la Torre~Luque}\ \emph {et~al.}(2023)\citenamefont
  {De~la Torre~Luque}, \citenamefont {Balaji},\ and\ \citenamefont
  {Carenza}}]{DelaTorreLuque:2023nhh}%
  \BibitemOpen
  \bibfield  {author} {\bibinfo {author} {\bibfnamefont {P.}~\bibnamefont
  {De~la Torre~Luque}}, \bibinfo {author} {\bibfnamefont {S.}~\bibnamefont
  {Balaji}}, \ and\ \bibinfo {author} {\bibfnamefont {P.}~\bibnamefont
  {Carenza}},\ }\href@noop {} {\  (\bibinfo {year} {2023})},\ \Eprint
  {http://arxiv.org/abs/2307.13728} {arXiv:2307.13728 [hep-ph]} \BibitemShut
  {NoStop}%
\bibitem [{\citenamefont {Cheek}\ \emph {et~al.}(2022)\citenamefont {Cheek},
  \citenamefont {Heurtier}, \citenamefont {Perez-Gonzalez},\ and\ \citenamefont
  {Turner}}]{Cheek:2021odj}%
  \BibitemOpen
  \bibfield  {author} {\bibinfo {author} {\bibfnamefont {A.}~\bibnamefont
  {Cheek}}, \bibinfo {author} {\bibfnamefont {L.}~\bibnamefont {Heurtier}},
  \bibinfo {author} {\bibfnamefont {Y.~F.}\ \bibnamefont {Perez-Gonzalez}}, \
  and\ \bibinfo {author} {\bibfnamefont {J.}~\bibnamefont {Turner}},\ }\href
  {\doibase 10.1103/PhysRevD.105.015022} {\bibfield  {journal} {\bibinfo
  {journal} {Phys. Rev. D}\ }\textbf {\bibinfo {volume} {105}},\ \bibinfo
  {pages} {015022} (\bibinfo {year} {2022})},\ \Eprint
  {http://arxiv.org/abs/2107.00013} {arXiv:2107.00013 [hep-ph]} \BibitemShut
  {NoStop}%
\end{thebibliography}%

\clearpage

\onecolumngrid
\begin{center}
  \textbf{\large Supplementary Material for Enhanced Constraints on Primordial Black Holes Using X-Ray Telescopes}\\[.2cm]
  \vspace{0.05in}
  { Jun Guo, Qiang Yuan, Bin Zhu}
\end{center}

\twocolumngrid
%%%%%%%%%% Merge with supplemental materials %%%%%%%%%%
\setcounter{equation}{0}
\setcounter{figure}{0}
\setcounter{table}{0}
\setcounter{section}{0}
\setcounter{page}{1}
\makeatletter
\renewcommand{\theequation}{S\arabic{equation}}
\renewcommand{\thefigure}{S\arabic{figure}}
\renewcommand{\thetable}{S\arabic{table}}

\onecolumngrid

This Supplementary Material (SM) is organized as follows.  

\section{Computational Framework for X-ray}
\label{supp:X-ray}

\subsection{Primordial Black Hole Hawking Evaporation}
In this section, we provide additional details and derivations to support the analysis presented in the main text. We begin by discussing the temperature of a non-rotating Black Hole (BH) and its relation to the emission of fundamental particles. We then introduce the expression for the particle emission rate, incorporating the greybody factors that depend on the particle species. Finally, we present the calculation of the differential photon flux per solid angle, accounting for the contributions from all particle species emitted during the BH evaporation process.

The temperature $T_{\mathrm{BH}}$ of a non-rotating BH with mass $M$ is given by Equation~\ref{Eq:temperature},
\begin{equation}
T_{\mathrm{BH}}=\frac{1}{8 \pi G_N M} \approx 1.06 \mathrm{MeV}\left(\frac{10^{16} \mathrm{gm}}{M}\right),
\label{Eq:temperature}
\end{equation}
where $G_N$ is the Newtonian gravitational constant. A BH of this kind emits fundamental particles of each species at a certain rate, as described by Equation~\ref{Eq:phbspec},
\begin{equation}
\frac{\partial^2 N_i}{\partial E_i \partial t}=\frac{g_i}{2 \pi} \frac{\Gamma_i\left(T_{\mathrm{BH}}, E_i, s_i\right)}{\exp \left(E_i / T_{\mathrm{BH}}\right)-(-1)^{2 s_i}},
\label{Eq:phbspec}
\end{equation}
where $i$ represents the particle species, $g_i$ denotes the degrees of freedom, $E_i$ is the energy of the emitted particle, and $s_i$ represents its spin. The greybody factors $\Gamma_i$ depend on the particle species $i$ and are related to the degrees of freedom $g_i$. Specifically, for photons ($g_{\gamma}=2$) and positrons ($g_e=2$), the greybody factors can be expressed as Equation~\ref{Eq:gamma},
\begin{equation}
\Gamma_i\left(E_i, T_{\mathrm{BH}}, s_i\right)=27\left(\frac{E_i}{8 \pi T_{\mathrm{BH}}}\right)^2 \gamma\left(E_i, T_{\mathrm{BH}}, s_i\right),
\label{Eq:gamma}
\end{equation}
where $\gamma$ is the factor obtained from Ref.~\cite{Cheek:2021odj}, {The greybody factor for a particle species $i$ can be derived from the wave equation near the black hole's event horizon. For simplicity, we consider the Schwarzschild metric, though the derivation can be extended to other metrics such as Kerr for rotating black holes. The equation governing the radial function $R(r)$ of a particle in the black hole's field is given by:
\begin{equation}
\left(\frac{d^2}{dr_*^2} + \omega^2 - V_{\mathrm{eff}}(r)\right)R(r) = 0,
\end{equation}
where $r_*$ is the tortoise coordinate, $\omega$ is the angular frequency of the particle, and $V_{\mathrm{eff}}(r)$ is the effective potential. The greybody factor is then given by the transmission coefficient $T(\omega)$ of the wave through the potential barrier:
\begin{equation}
\Gamma_i(E_i, T_{\mathrm{BH}}, s_i) = T(\omega) = \frac{R(r \rightarrow \infty)}{R(r \rightarrow r_{BH})},
\end{equation}
where $r_{BH}$ is the radius of the event horizon.}

\subsection{Differential Photon Flux Calculation}

{We now present a detailed step-by-step explanation of the calculation of the differential photon flux per solid angle. This calculation is pivotal to understanding the observable signatures of black hole evaporation.}

{The differential photon flux from a point-like source is given by:
\begin{equation}
\frac{\mathrm{d} N_\gamma}{\mathrm{d} A \mathrm{d} t \mathrm{d} \Omega \mathrm{d} E} = \frac{1}{4 \pi} \frac{\mathrm{d} \Phi^{\mathrm{PBH}}_\gamma}{\mathrm{d} E},
\end{equation}
where $A$ is the detector area, $t$ is the observation time, $\Omega$ is the solid angle, and $E$ is the photon energy. The total flux from an extended source, such as a black hole distribution, requires integration over the line of sight (LOS) and summation over all contributing black holes. The line-of-sight integral can be expressed as:
\begin{equation}
\int_{\mathrm{LOS}} dl \frac{\partial^2 N_\gamma}{\partial E_\gamma \partial t} = \int_0^{D} dr \, n_{\mathrm{BH}}(r) \frac{\partial^2 N_\gamma}{\partial E_\gamma \partial t},
\end{equation}
where $n_{\mathrm{BH}}(r)$ is the number density of black holes at a distance $r$ from the observer, and $D$ is the maximum distance considered for the integration.}

To calculate the differential photon flux per solid angle from a region defined by the angular direction, we integrate the photon yield $N_{\gamma}$ overall particle species emitted by the BH during its evaporation process. This is expressed as Equation~\ref{Eq:flux}:
\begin{equation}
\frac{\mathrm{d} \Phi^{\mathrm{PBH}}\gamma}{\mathrm{d} E\gamma}=\frac{1}{4 \pi} \int_{\mathrm{LOS}} d l \frac{\partial^2 N_\gamma}{\partial E_{\gamma} \partial t} \frac{f_{\mathrm{PBH}} \rho_{\mathrm{DM}}(l, \psi)}{M},
\label{Eq:flux}
\end{equation}
where $\frac{\partial^2 N_\gamma}{\partial E_\gamma \partial t}$ represents the total photon emission rate. It includes contributions from primary and secondary photons resulting from the decay and subsequent interactions of emitted particles. These contributions can be expressed as Equation~\ref{Eq:spectra}:
\begin{equation}
\begin{aligned}
\frac{\partial^2 N_\gamma}{\partial E_\gamma \partial t}&=\frac{\partial^2 N_{\gamma, \text { primary }}}{\partial E_\gamma \partial t}\
&+\sum_{i=e^{ \pm}, \mu^{ \pm}, \pi^{ \pm}} \int \mathrm{d} E_i \frac{\partial^2 N_{i, \text { primary }}}{\partial E_i \partial t} \frac{\mathrm{d} N_i^{\mathrm{FSR}}}{\mathrm{d} E_\gamma} \
&+\sum_{i=\mu^{ \pm}, \pi^0, \pi^{ \pm}} \int \mathrm{d} E_i \frac{\partial^2 N_{i, \text { primary }}}{\partial E_i \partial t} \frac{\mathrm{d} N_i^{\text {decay }}}{\mathrm{d} E_\gamma}.
\label{Eq:spectra}
\end{aligned}
\end{equation}

\subsection{X-ray from the Inverse Compton Scattering}
In the previous section, we discussed the computation of the photon spectrum resulting from primary black hole (PBH) evaporation. However, that computation did not account for the photon spectra generated by inverse Compton scattering (ICS). Inverse Compton scattering refers to the interaction between high-energy electrons (muons/pions) and photons, where the electrons gain energy from the photons through scattering.

The differential flux of inverse Compton scattering emission, denoted as ${d \Phi_{\gamma}^{\mathrm{IC}}}/{d E_\gamma d \Omega}$, can be represented by the following equation:
\begin{equation}
\frac{d \Phi_{\gamma}^{\mathrm{IC}}}{d E_\gamma d \Omega}=\frac{1}{E_\gamma} \int_{\text {l.o.s. }} d s \frac{j\left(E_\gamma, s, b, \ell\right)}{4 \pi}.
\label{Eq:icsEmission}
\end{equation}

In equation~\ref{Eq:icsEmission}, $E_\gamma$ represents the energy of the emitted photons, and the integral over the line of sight (l.o.s.) accounts for the contributions from different locations along the line of sight. The term $j\left(E_\gamma, s, b, \ell\right)$ represents the emissivity, which is obtained by convolving the power of inverse Compton scattering ($\mathcal{P}_{\mathrm{IC}}$) with the differential number density ($dn{e^{\pm}}/dE_e$) of the electrons and positrons that emit radiation at that point. The power of inverse Compton scattering, denoted as $\mathcal{P}_{\mathrm{IC}}$, is given by the following equation:
\begin{equation}
\mathcal{P}_{\mathrm{IC}}\left(E\gamma, E_e, \textbf{r}\right)=E_\gamma \int d \epsilon n_\gamma(\epsilon, \textbf{r}) \sigma_{\mathrm{IC}}\left(\epsilon, E_\gamma, E_e\right).
\label{Eq:PIC}
\end{equation}
Here, $E_e$ represents the energy of the electrons, $\textbf{r}$ represents the position, $n_\gamma(\epsilon, \textbf{r})$ represents the photon number density, and $\sigma_{\mathrm{IC}}$ is the cross-section for inverse Compton scattering. The analytical formula for $\sigma_{\mathrm{IC}}$ can be found in the PPPC (Particle Physics Phenomenology Calculator) package, specifically in the references Cirelli et al. (2010), Cirelli et al. (2009), and Meade et al. (2009).

The differential number density of electrons and positrons, denoted as ${d n_{e^{ \pm}}}/{d E_e}(E_e, \textbf{r})$, is given by
\begin{equation}
\frac{d n_{e^{ \pm}}}{d E_e}(E_e, \textbf{r})=\frac{f_{\mathrm{PBH}}}{b_{\text {tot }}(E_e, \textbf{r})}\int_{E_e}^{m_{\mathrm{DM}} / 2} d \tilde{E}e \frac{\rho(\textbf{r})}{M}\frac{\partial^2 N{e^{ \pm}}}{\partial \tilde{E}_e\partial t},
\label{Eq:dnedEe}
\end{equation}
where $f_{\mathrm{PBH}}$ represents the fraction of dark matter in the form of PBHs, $b_{\text{tot}}$ is the energy loss function (which can be found in the PPPC package), $m_{\mathrm{DM}}$ represents the mass of the dark matter particle, $\rho(\textbf{r})$ represents the dark matter density at a specific position, and $\frac{\partial^2 N_{e^{ \pm}}}{\partial \tilde{E}_e\partial t}$ represents the double differential number flux of electrons and positrons.

Taking into account the photon spectra resulting from inverse Compton scattering, the total photon spectrum can be expressed as
\begin{equation}
\frac{\mathrm{d} \Phi_\gamma^{\mathrm{tot}}}{\mathrm{d} E_\gamma}=\frac{\mathrm{d} \Phi^{\mathrm{PBH}}\gamma}{\mathrm{d} E\gamma}+\frac{d \Phi_{\gamma}^{\mathrm{IC}}}{d E_\gamma}.
\end{equation}
This equation combines the photon spectrum from primary black hole evaporation ($\frac{\mathrm{d}\Phi^{\mathrm{PBH}}\gamma}{\mathrm{d} E\gamma}$) with the photon spectrum from inverse Compton scattering ($\frac{d \Phi_{\gamma}^{\mathrm{IC}}}{d E_\gamma}$) to obtain the total photon spectrum.

\section{Spectra}
\begin{figure}
\begin{center}
\includegraphics[scale=0.4]{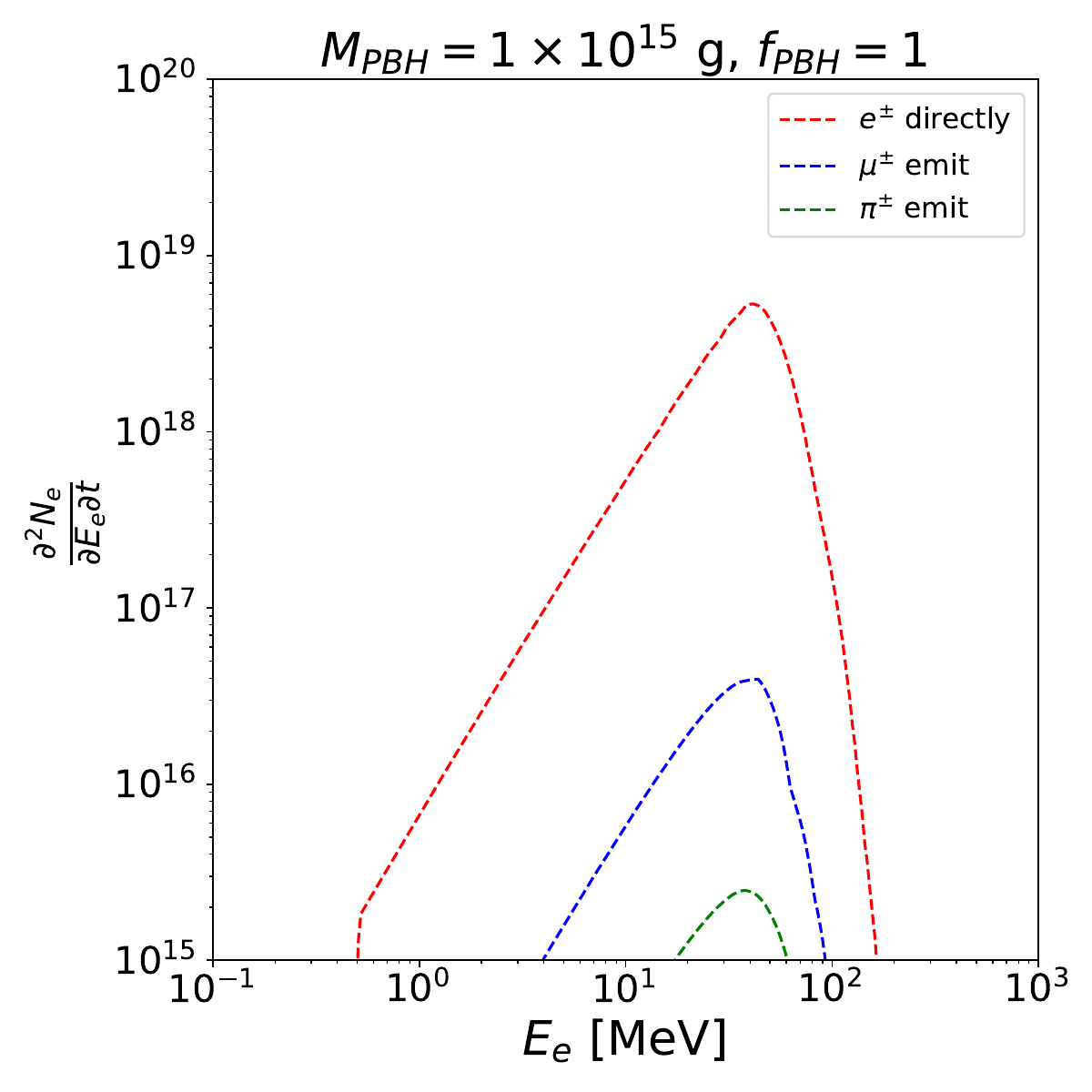}
\includegraphics[scale=0.4]{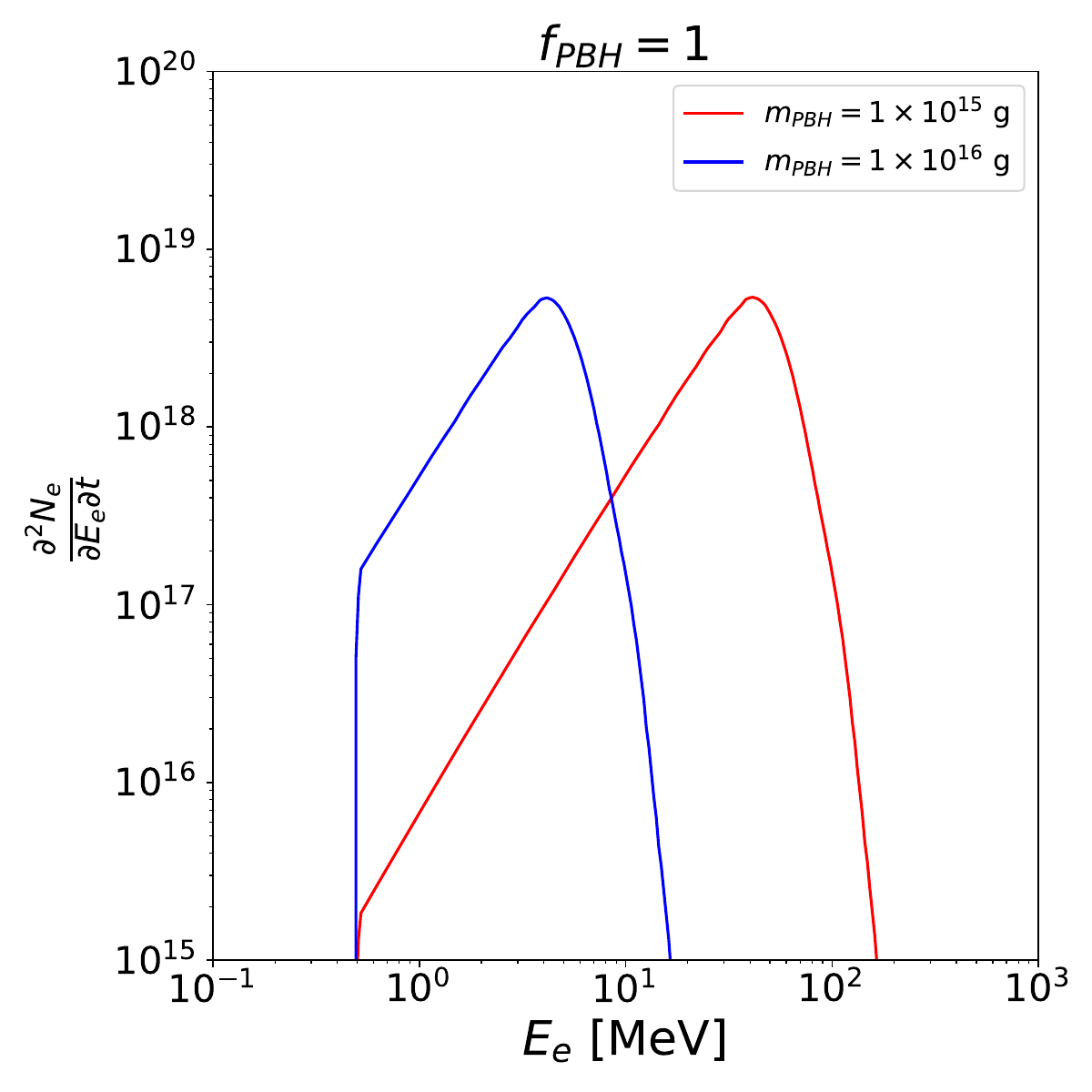}
\end{center}
\caption{Left: the three dominant positron generating channel, $e^\pm$ direct emission makes the dominant
contribution. Right: small mass PBH with $m_{\rm PBH} = 1\times 10^{15}$ g will produce a sharp positron emission, for a large mass PHB, the positron emission is mainly concentrate on low energy region. }
\label{positron_spec}
\end{figure}

The ICS photon is excited by final-state electrons. For the PBH case,  the dominant $e^{\pm}$ final products come from three 
channels,  the PBH direct generate $e^{\pm}$, the PBH radiates $\mu^{\pm}$ and $\pi^{\pm}$ first, and 
then the radiational products decay into $e^{\pm}$. The direct emitted $e^{\pm}$  can be calculate 
according to Eq.~\ref{Eq:phbspec}, and for the $e^{\pm}$ from radiated $\mu^{\pm}$ and $\pi^{\pm}$ decay, they could be given as
\begin{equation}
\frac{\partial^2N_e}{\partial E_e \partial t} = \int dE' \frac{\partial^2N_{\mu(\pi)}}{\partial E'_{\mu(\pi)} \partial t} \frac{dN_e}{dE}(E'),
\end{equation}
where $\frac{dN_e}{dE}(E')$ is the $\mu$($\pi$) decay positron spectrum with initial energy $E'$, given by boosting the 
decay positron spectrum obtained in the mother particle rest frame:
\begin{equation}
\frac{dN}{dE} = \frac{1}{2\beta\gamma}\int^{E'_{max}}_{E'_{min}} \frac{1}{p'}\frac{dN}{dE'},
\end{equation}
where $\gamma = E_A / m_A$ and $\beta = \sqrt{1-1/\gamma^2}$ are boost factors. We generate the positron spectrum by using Hazma~\cite{} and show the result for the case $m_{\rm PBH} = 1\times 10^{15}$ g and $f_{\rm PBH} = 1$ on left panel in Fig.~\ref{positron_spec}, it is clear that the direct emitted $e^{\pm}$
dominant positron production, this is because to generate a positron according to radiated
muon or poin, the least energy of pion and muon radiation should be $m_{\mu(\pi)}$, at same 
time, muon and pion masses is about 100 MeV, which receive a great suppression from numerator in Eq.~\ref{Eq:phbspec}, accordingly, for a heavier PHB with $m_{\rm PBH} = 1\times 10^{16}$ g, the muon 
and pion radiation will contribute little to the positron emission. 

On the right panel of 
Fig.~\ref{positron_spec}, we show the total positron emission in the case of $m_{\rm PBH} = 1\times 10^{15}$ g and $m_{PBH} = 1\times 10^{16}$ g, it is clear to see PBH with lighter mass will generate
sharper positron spectrum.

\section{Model Spectrum Evaluation}
The spectrum of a DM source is described by its photon spectrum $f(E)$, which is a continuous functon in the unit of photon $\rm m^{-2}\rm s^{-1}\rm keV^{-1}$, but the signal observed by the experimental instruments $s(E)$ is different from $f(E)$, they are connected by the instrument response function $R(E, E')$ in a convolution process:
\begin{align}
s(E) = \int f(E') R(E, E')dE'.
\label{sig_conv}
\end{align}
The $R(E, E')$ plays the role of effective area and is in the unit of $m^2$, $s(E)$ means the observed data in the energy channel $E$, and the unit of which is count~$\rm s^{-1}~\rm keV^{-1}$. But in the reality experiments, the energy of certain channel is not an arbitrary value, but some limited number of energy bins, so the detected signal should be considered  belonging to certain bin and the response function should be the connections between different bins. The response function is replaced by a discrete response matrix $R_{ij}$, and the spectrum function should be replaced by 
\begin{align}
    F_i = \int^{E_{ui}}_{E_{di}}f(E)dE\simeq f\left(\frac{E_{ui} + E_{di}}{2}\right)(E_{ui} - E_{di}).
\end{align}
The instrument expected observe value could be carry out as:
\begin{align}
S_i = \sum_j R_{ij}F_j,
\label{response_matrix}
\end{align}
where $i$ is the bin index observed by instrument and $j$ is the bin index of model spectrum.

\end{document}